 \documentstyle[aps,twocolumn,epsf]{revtex}

\begin{document}

\draft

 \twocolumn[\hsize\textwidth\columnwidth\hsize\csname@twocolumnfalse\endcsname

\title{Symmetry Does Not Allow Canting of Spins in La$_{1.4}$Sr$_{1.6}$Mn$_2$O$_7$}
\author{Fan Zhong$^{1,2}$ and Z. D. Wang$^1$}
\address{$^1$Department of Physics, University of Hong Kong, Hong Kong, People's Republic of China\\
$^2$Department of Physics, Zhongshan University, Guangzhou 510275, People's Republic of China\cite{byline}}
\date{October 28, 1999}
\maketitle

\begin{abstract}
We analyze the symmetry of all possible magnetic structures of bilayered
manganites La$_{2-2x}$Sr$_{1+2x}$Mn$_2$O$_7$ with doping $0.3 \leq x < 0.5$
and formulate a corresponding Landau theory of the phase transitions
involved. It is shown that canting of spins is not allowed at $x=0.3$
though is at $x=0.4$. The observed magnetic reflections from the sample with
$x=0.3$ may be described as arising from two spatially distributed
phases with close transition temperatures but different easy axes and ranges
of stability. Experimental results are revisited on the basis of the theoretical findings.
\end{abstract}

\vspace{0.2in}


\preprint{ZHONG Fan}
 ]


Recent extensive investigation of the so-called colossal
magnetoresistance (CMR) \cite{cmr} in doped perovskite manganites has
stimulated considerable interest in relative bilayered compound
La$_{2-2x}$Sr$_{1+2x}$Mn$_2$O$_7$ in an attempt to understand and to improve
the sensitivity of the magnetoresistive response \cite{moritomo96,kimura96,kimura97}.
The material of interest is comprised of perovskite (La, Sr)MnO$_3$ bilayers
separating by (La, Sr)O blocking layers, namely, the $n=2$ member of the
Ruddlesden-Popper series of manganites (La, Sr)O[(La, Sr)MnO$_3$]$_n$.
This quasi two-dimensional nature promotes fluctuations that lower the
critical temperature $T_c$ of the magnetic transition and hence the relevant
scale of a magnetic field for the huge magnetoresistance. As the tetragonal
$I4/mmm$ symmetry of the material {\em a priori} lifts the degeneracy of
the $e_g$ orbitals of the Mn$^{3+}$ ions, the Jahn-Teller distortion of
which was argued to be responsible for the CMR of the perovskite manganites
\cite{millis}, observation of antiferromagnetic (AFM) correlations
above $T_c$ of a para- (PM) to ferromagnetic (FM) transition in
La$_{1.2}$Sr$_{1.8}$Mn$_2$O$_7$ was suggestive as an alternative origin to
assist localization of carriers above $T_c$ \cite{perring97}. Importance of
the AFM superexchange interaction shows up at the same doping level as
canting of the ordered moments in neighboring layers within each bilayer as
inferred from the sign reversal of the Mn-O bond compressibility below
$T_c$ \cite{argyriou97}. Further neutron scattering investigation of PM
correlations provided evidence for the strong canting of the spins with
an average angle that depends on both the magnetic field and the temperature
above $T_c$ owing to the weaker FM correlation within the bilayers \cite{osborn}.
The canting angle, in particular, changes from $86^\circ$ at zero field to
$74^\circ$ at an external magnetic field of 1 Tesla to $53^\circ$ at 2 Teslas
at 125K. Comprehensive neutron-diffraction studies on the other hand found
that the canting angle increases from 6.3$^{\circ}$ at $x=0.4$ to
180$^{\circ}$ (A-type AFM) at $x=0.48$ at 10K, while $T_c$ decreases from
120 K to 0 K correspondingly. Moreover, the AFM correlations above $T_c$
were identified as an intermediate phase whose order parameter decreases
in an anomalous exponential manner upon increasing
temperature to about 200K \cite{hirota}. Accordingly, the AFM correlations
and more generally the magnetic structure seem to play an important role in
the bilayered manganites.

For $0.32 \alt x \alt 0.4$, the bilayered manganites exhibit
a FM order below $T_c$ with an easy axis at the layer. The magnetic
structure at $x=0.3$, however, is somewhat complicated and so there
exists no consensus. Perring {\it et al} \cite{perring98} proposed an AFM
order of an intra-bilayer FM and inter-bilayer AFM structure (denoted as
AFM-B) with the easy axis along $z$ below about 90K from magnetic neutron
diffraction. However, a substantial component within the layers
rises up and then falls down between 60 and 90K or so. Argyriou {\it et al}
\cite{argyriou99} by neutron diffractions and Heffner {\it et al} \cite{heffner}
by muon spin rotation measurements reported, 
on the other hand, that their sample with the same
doping involves two structurally similar phases: The major phase (hole poor)
arranges itself in a similar AFM-B structure with a substantial canting in
the plane as well as out of it. The minor phase (hole rich but $x<0.32$)
differs from the major one only by its FM arrangement along $z$ axis
and its lower ordering temperature. However, as they pointed out, the
assignment of the in-plane component is not so unambiguous. Also their
in-plane AFM reflections become vanishingly small below about 60K either. 
Still another scenario at the 30 percent doping is this: The magnetic
structure changes from PM to AFM-B at about 100K and then to FM at 70K or
so. The easy axis rotates correspondingly from in-plane in the AFM-B to $z$
direction in the FM state \cite{kimura97,moritomo99,kubota}. From these
experiments, whether there exists canting of spins at $x=0.3$ is still ambiguous.
So, noticing the importance of the magnetic structure in the
$x \agt 0.4$ doping, clarification of the magnetic structure of the $x=0.3$
doping is a key to understand its characteristic transport
behavior \cite{kimura96,li}. In this Letter, we show that there is a
qualitative difference between doping at $x=0.3$ and $x=0.4$ by analyzing the
symmetry of the magnetic structures. It is found that the symmetry of the
magnetic order parameters cannot allow canting at $x=0.3$ in contrast to 
$x=0.4$. This result sheds new light to the mechanism of the CMR behavior.

\begin{figure}
\epsfysize 2.3in \epsfbox{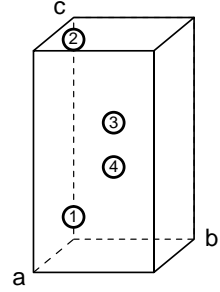}
\caption{Elementary unit cell of $I4/mmm$ with four Mn ions and their numbering.}
\label{cell}
\end{figure}

First we identify the order parameters and their sym-
\begin{minipage}{20.5pc}
\begin{table}
\caption{Components of the magnetic vectors that form a basis of the IR's of $I4/mmm$ at ${\bf k}_{\Gamma}$ and ${\bf k}_M$.}
\label{bases}
\begin{center}
\begin{minipage}{11pc}
\begin{tabular}{cc}
IR & BASES \\ \tableline
$\tau ^{2}$ & $L_{z};\ L_{Az}$ \\
$\tau ^{3}$ & $M_{z};\ L_{Bz}$ \\
$\tau ^{9}$ & $(M_{x},M_{y});\ (L_{Bx},L_{By})$ \\
$\tau ^{10}$ & $(L_{x},L_{y});\ (L_{Ax},L_{Ay})$
\end{tabular}
\end{minipage}
\end{center}
\end{table}
\end{minipage}
metry responsible for
the possible magnetic structures. The Mn ions with magnetic moments
{\boldmath $\mu$}$_{i}$ in the $I4/mmm$ structure occupy four
positions at $i=1(0,0,z)$, $2(0,0,1-z)\ (z \sim 0.1)$ and their translation by
${\bf t}_0=(\frac{1}{2},\frac{1}{2},\frac{1}{2})$,
i.e., $(\frac{1}{2},\frac{1}{2},\frac{1}{2}\pm z)$ (see Fig.~\ref{cell})\cite{inter}.
Following the representation analysis of magnetic structures
\cite{dzyal,toledano,zhong99}, we define two magnetic vectors
\begin{eqnarray}
{\bf M} & = & \mbox{{\boldmath $\mu$}}_{1}+\mbox{{\boldmath $\mu$}}_{2}
\nonumber \\
{\bf L} & = & \mbox{{\boldmath $\mu$}}_{1}-\mbox{{\boldmath $\mu$}}_{2}.
\end{eqnarray}
Then a FM state corresponds to {\bf M} propagating with a wave vectors
${\bf k}_{\Gamma}=(000)$, a bilayered-type AFM-B and an A-type AFM
(intra-bilayer AFM but inter-bilayer FM) state to {\bf M} and {\bf L},
respectively, with ${\bf k}_M=(00\frac{1}{2})$ of the first Brillouin zone.
Denoting the latter two order parameters as ${\bf L}_B$ and ${\bf L}_A$
respectively, and noticing that {\bf k}$_{\Gamma}$ and {\bf k}$_M$ share the same
irreducible representations (IR's) of the $I4/mmm$ group \cite{kovalev},
one can find the components of the four vectors that form bases of the IR's
shown in Table \ref{bases}.
Note that the IR's $\tau^9$ and $\tau^{10}$ are both two-dimensional, and so
$M_{x}$ and $M_{y}$ together form a basis vector of $\tau^9$, so do $L_{Bx}$
and $L_{By}$. From Table \ref{bases} and the possible experimental magnetic
structures \cite{hirota,argyriou99,kubota}, we identify {\bf L}$_{B}$ with
the order parameter for the major phase, $M_z$ and ($L_{Bx}, L_{By}$) for
the minor phase of $x=0.3$, $(M_x, M_y)$ with the order parameter for
$0.3<x \alt 0.38$, $(M_x, M_y)$ and $(L_{Ax}, L_{Ay})$ for $0.38 <x<0.48$,
and $(L_{Ax}, L_{Ay})$ for $ 0.48 \alt x<0.5$ which is A-type AFM.

From Table \ref{bases}, the relevant lowest order magnetic part of the
Landau free-energy can be written as
\begin{eqnarray}
F & = & \frac{c}{2}{\bf M}^{2} + \sum_{w}\frac{a_{w}}{2}{\bf L}_{w}^{2}
+\sum_{w}\frac{b_{w}}{4}{\bf L}_{w}^{4}+\frac{d}{4}{\bf M}^{4} \nonumber \\
& & + \frac{1}{2}\beta _{z}M_z^2 + \frac{1}{2}\beta _{xy}(M_x^2+M_y^2) \nonumber \\
& &  + \sum_{w}\left[ \frac{1}{2}\alpha_{wxy}(L_{wx}^2+L_{wy}^2)+ \frac{1}{2}\alpha_{wz}L_{wz}^2 \right], \label{fall}
\end{eqnarray}
where $w$ represents the summation over ${\bf L}$, ${\bf L}_A$,
and ${\bf L}_B$. Note that the latter two vectors will carrier a factor
$\exp\{-i{\bf k}_M \cdot {\bf t}_0\}=-1$ when they are translated by ${\bf t}_0$, and
so cannot appear in odd powers. In Eq.~(\ref{fall}), we have separated the
exchange contributions (first four terms), which depend only on the relative
orientation of the spins, from the magnetic anisotropic energies (remaining
terms), which depend on the relative direction of the magnetic moments to
the lattice and arise from the rela-
\noindent
tivistic spin-spin and spin-orbit
interactions and so are effects of the order of $O(v_0^2/c_0^2)$, ordinarily about
$10^{-2}$ to $10^{-5}$, where $v_0$
is the speed of electrons in the crystal and $c_0$ that of light, since the
magnetic moments themselves contain a factor $v_0/c_0$ \cite{landau}.
Hence $\alpha$ and $\beta$ are small constants due to their relativistic origin.
$b_{w}$ and $d$ are positive for stability.

We now focus on the $x=0.3$ doping. The relevant magnetic vectors in this
case is ${\bf L}_B$ and ${\bf M}$. Minimizing Eq.~(\ref{fall}) with the
components of these vectors, one obtains five solutions
\begin{mathletters}
\label{phase3}
\begin{eqnarray}
{\bf M}={\bf L}_B={\bf 0}, \label{phase0} \\
{\bf M}={\bf 0}, L_{Bx}=L_{By}=0, L_{Bz}^2=-\frac{a_B+\alpha_{Bz}}{b}, \label{phaselz} \\
{\bf M}={\bf 0}, L_{Bz}=0, L_{Bx}^2+L_{By}^2=-\frac{a_B+\alpha_{Bxy}}{b}, \label{phaselxy} \\
{\bf L}_B={\bf 0}, M_x=M_y=0, M_z^2=-\frac{c+\beta_z}{d}, \label{phasemz} \\
{\bf L}_B={\bf 0}, M_z=0, M_x^2+M_y^2=-\frac{c+\beta_{xy}}{d}. \label{phasemxy}
\end{eqnarray}
\end{mathletters}
Since anisotropic terms like $M_x^2M_y^2$ have not been included, the
direction in the $xy$-plane cannot yet be determined. Note that the exchange
term of $({\bf L}_B \cdot {\bf M})^2$ type is irrelevant, since
${\bf M}\cdot {\bf L}_B=0$ due to the incompatibility of ${\bf M}$ and
${\bf L}_B$ along a single direction. Eq.~(\ref{phase0}) represents the PM
phase, Eqs.~(\ref{phaselz}) and (\ref{phaselxy}) pure AFM-B phases with the
moments directing respectively along the $z$-axis and the $xy$-plane, and
Eqs.~(\ref{phasemz}) and (\ref{phasemxy}) pure FM phases. An remarkable
feature of Eqs.~(\ref{phase3}) is that there is no mixed order such as
$L_{Bz}$ with $L_{Bx}$ or $L_{By}$, $M_z$ with $M_x$ or $M_y$ and
${\bf L}_B$ with ${\bf M}$. In other words, no canting state exists. 
The reason is that there is no symmetry relation
between $\alpha_{Bz}$ ($\beta_z$) and $\alpha_{Bxy}$ ($\beta_{xy}$), so that
both $L_{Bz}$ ($M_z$) and $L_{Bx}$ ($M_x$) or $L_{By}$ ($M_y$) cannot
simultaneously acquire nonzero values in general. This can also been seen
from Table \ref{bases} that the $z$ and the $xy$ components transform
according to different IR's.

In order to determine the range of stability of the phases, one substitutes
the solutions Eqs.~(\ref{phase3}) into the free energy and obtains
respectively to the first order,
\begin{mathletters}
\label{phasefe}
\begin{eqnarray}
F_0 & = & 0, \label{fphase0}\\
F_{L_z} & \simeq & -\frac{a_B^2}{4 b_B}-\frac{a_B \alpha_{Bz}}{2 b_B}, \label{fphaselz} \\
F_{L_{xy}} & \simeq & -\frac{a_B^2}{4 b_B}-\frac{a_B \alpha_{Bxy}}{2 b_B}, \label{fphaselxy} \\
F_{M_z} & \simeq & -\frac{c^2}{4 d}-\frac{c \beta_{z}}{2 d}, \label{fphasemz} \\
F_{M_{xy}} & \simeq & -\frac{c^2}{4 d}-\frac{c \beta_{xy}}{2 d}. \label{fphasemxy} \end{eqnarray}
\end{mathletters}
Accordingly, if $0<\beta_z<\beta_{xy}$, for instance, then $F_{M_z}<F_{M_{xy}}$ and so the moments will point to $z$-axis, whereas, if $\beta_z > \beta_{xy}>0$, they will lie on the $xy$-plane. This may be 
the case for the change of the FM magnetization direction with increasing
doping observed experimentally \cite{kubota}. Similarly, when
$\alpha_{Bz}$ becomes bigger than $\alpha_{Bxy}$ (both are assumed to be
positive without loss of generality), the system changes from the phase
$L_{Bz}$ [Eq.~(\ref{phaselz})] to $L_{Bxy}$ [Eq.~(\ref{phaselxy})]. The two
phases have respectively crystallographic space groups $P4/mnc$ and $Cmca$,
which cannot be related by an active IR and so the transition between them
is necessarily discontinuous \cite{stokes}. Another reason is that the two
directions are not connected continuously. In practice, the two phases may
appear almost simultaneously within of a single sample at different places
where there is, for example, a small variation of doping or inhomogeneity
since the two phases differ in their transition points [$a_B+\alpha_B=0$,
Eqs.~(\ref{phase3})] and free energies by only values of the order of
$O(v_0^2/c_0^2)$, and so which will appear depend rather sensitively on
detailed conditions. This same reason also implies that the separation might
be mesoscopic. Moreover, the two phases may have different temperature
windows of stability due to different variations of $\alpha_{Bz}$ and
$\alpha_{Bxy}$ with the temperature. Occurrence of AFM-B or FM order relies
on the other hand on whether $a_B$ or $c$ becomes negative first, respectively.

There exists possible mixing of $L_{Bz}$ and its $xy$-plane counterparts at
higher order terms, but it cannot produce canting either. As the
transition points of the two phases differ by only small quantities of order
of $O(v_0^2/c_0^2)$, we use the expansion in ${\bf L}_B$ itself. Thus,
besides those pure ${\bf L}_B$ terms in the free energy Eq.~(\ref{fall}),
we add terms
\begin{eqnarray}
& & \frac{1}{4}\lambda_{1}L_{Bz}^4,\ \frac{1}{4}\lambda_{2}(L_{Bx}^2+L_{By}^2)^2,\nonumber \\
& & \frac{1}{2}\lambda_{3}L_{Bz}^2 (L_{Bx}^2+L_{By}^2),\ \frac{1}{2}\lambda_{4}L_{Bx}^2 L_{By}^2,
\end{eqnarray}
with the coefficients $\lambda$'s of order $O(v_0^4/c_0^4)$ relative to the
exchange ones \cite{dzyal}. Then one can obtain new solutions that determine the
direction of the moments in the $xy$-plane to be either along the $x$ or $y$
axis or along its diagonal depending respectively on whether $\lambda_4$ is
positive or negative. In addition, there appear solutions such as
\begin{mathletters}
\begin{eqnarray}
L_{Bx} & = & 0,\\
L_{By}^2 & \simeq & -\frac{a_B (\lambda_1-\lambda_4)+b_B (\alpha_1-\alpha_2)}{b_B (2\lambda_1-\lambda_3-\lambda_4)}, \label{pseudox} \\
L_{Bz}^2 & \simeq & -\frac{a_B (\lambda_1-\lambda_3)-b_B (\alpha_1-\alpha_2)}{b_B (2\lambda_1-\lambda_3-\lambda_4)}, \label{pseudoz}
\end{eqnarray}
\end{mathletters}
and a similar one in the diagonal plane perpendicular to the $xy$-plane,
where we have kept terms of order $\lambda$ in both the numerators and
denominators. However, it is readily seen that the left hand sides of
Eqs.~(\ref{pseudox}) and (\ref{pseudoz}) possess just opposite signs in
general, so that only one of them can have a real solution. Similar result
can also be proved by expanding the free energy in the unit vector along
${\bf L}_B$ valid at low temperatures. Further, there is no external or
demagnetizing field to tilt the moments.
Therefore, canting is not allowed for the bilayered-type AFM order of the
major phase with $x=0.3$ doping. The observation of both the $z$ and the
$xy$ components of the AFM-B order should thus arise from the two phases
each with one kind of the AFM-B components.

Nevertheless, mixing of different magnetic vectors is still possible by
coupling of the type ${\bf M}^2{\bf L}_B^2$ for instance. This can exist
due to either an exchange or a relativistic origin. Adding such a term with
a coefficient $\delta/2$ for the coupling of, say, $M_z$ and $L_{Bx}$ and
$L_{By}$ for the minor phase of $x=0.3$, one obtains, besides
Eqs.~(\ref{phasemz}) and (\ref{phaselxy}), a new phase with mixing
\begin{eqnarray}
M_z^2 & \simeq & \frac{\delta a_B - c b_B}{d b_B -\delta^2}, \nonumber\\
L_{Bx}^2+L_{By}^2 & \simeq & \frac{\delta c- d a_B}{d b_B -\delta^2},
\label{phaseml}
\end{eqnarray}
where we have neglected $\alpha_B$ and $\beta$. A system
with such a coupling may exhibit several scenarios depending on the strength
of the coupling and the nature of the pure phases \cite{imry,toledano}. It
may appear in a pure phase, which may transform continuously or
discontinuously to the mixed phase, or discontinuously to another pure phase
at lower temperatures, the latter can only take place in the strong coupling
of $\delta^2> d b_B $. It may even change directly to the mixed phase when
the transition temperatures of the two pure phases get identical. Reentrant
phase transitions from a pure phase to a mixed one and then back to the pure
phase are also possible.

We now compare our results with experiments. The experimental assignment of
both a canting major phase and a canting minor phase is based on the result
that if {\em canting} is exclusively associated with only one phase, the
resultant total magnetic moment is too large at 80K, near the peak
temperature of the plane AFM reflections \cite{argyriou99}. This excludes
the possibility of a canting minor phase {\em and} a pure $L_{Bz}$ phase and
appears to suggest instead that the plane AFM reflections arise at least
partly from an independent $L_{Bxy}$ phase. The fact that the reflections
from $L_{Bz}$ and $L_{Bxy}$ start appearing at
almost the same temperature seems to support the theoretical results that
both phases emerge almost simultaneously at different places where there is
a small variation of doping or inhomogeneity, which balances the small
quantities $\alpha_{Bz}$ and $\alpha_{Bxy}$ in their transition
temperatures. With the two phases rather than a single canting major phase,
the too large magnetic moment may be remedied. The peak structure of the
reflection intensities from the $L_{Bxy}$ phase \cite{perring98,argyriou99}
may then arise from the different temperature dependence of $\alpha_{Bxy}$
and $\alpha_{Bz}$ in such a way that below about 60K,
$\alpha_{Bxy} > \alpha_{Bz}$, and so the $L_{Bxy}$ phase transforms to the
$L_{Bz}$ phase by a reorientation transition. The small remaining
reflections may originate from the remnant $L_{Bxy}$ phase due to possible
inhomogeneity or supercooling.

For higher doping, noting that the reflections from the $M_z$ component
emerge separately and accompany with the decline of the $L_{Bxy}$
reflections \cite{argyriou99}, it seems that the minor phase may be a pure
FM phase with the $z$-axis as its easy orientation. Its significantly lower
$T_c$ of about 80K than those of slightly higher doping
\cite{argyriou99,kimura98} might result from its competition with the
$L_{Bxy}$ phase, which suppresses its occurrence via a positive $\delta$.
Nevertheless, a canting minor phase may still be possible, but its lower
$T_c$ and the peak feature of the $L_{Bxy}$ reflections should be properly
accounted for. When doping increases, $T_c$ increases but $\beta_{xy}$
becomes smaller than $\beta_z$, and so the moment aligns ferromagnetically
in the $xy$-plane. At high doping near 0.5, the A-type AFM
is the most stable state. In between these two cases, the two types of
states compete with each other via mixing terms similar to
Eqs.~(\ref{phaseml}), leading possibly to the lowering of their respective
transition temperatures \cite{osborn} and a $(M_x, M_y)$ and
$(L_{Ax}, L_{Ay})$ tilt as observed experimentally. The exponential-like
growth of the A-type AFM with cooling might be due to two-dimensional FM fluctuations.

In conclusion, noticing the importance of magnetic correlations to
magnetoresistive response, we have analyzed the symmetry of all possible
magnetic structures of bilayered manganites with doping $0.3 \leq x < 0.5$
on the basis of experimental results. A corresponding Landau theory of the
phase transitions involved is formulated. A prominent result is that the
ordered magnetic moments of the $x=0.3$ doping (the major phase
\cite{argyriou99}) cannot be canting though $x=0.4$ can, since the former
is characterized by a single magnetic vector ${\bf L}_B$ whereas the latter
by two different magnetic vectors, which may be mixed by an exchange or
relativistic mechanism. Such a result indicates that the magnetic structure
of the $x=0.3$ doping is far more complex than what has been proposed and
demands further experimental clarifications. Instead of a canting major
phase, there exist two spatially distributed phases with close transition
temperatures but different easy axes and ranges of stability, to which the
observed magnetic reflections from the $x=0.3$ sample may be attributed. Such a picture
can account for the peak of the plane AFM reflections. Furthermore, it
seems to accord with the two-step variation of lattice parameters with
temperatures through an assumption that the $d_{3z^2-r^2}$ and
$d_{x^2-y^2}$ orbital states correspond to magnetic orientations along $z$
and $xy$ respectively, namely, an increase in the $L_{Bxy}$ phase elongates
the in-plane scale but shortens the $z$ scale, and then a decrease gives
rise to a reverse effect \cite{argyriou99,kimura98}. As both the $z$ and
the $xy$ components possess a bilayered-type AFM structure, the material
should be expected to display an insulating behavior in the whole
temperature range. So the metal-insulator transition should mostly be
attributed to the percolation of the minor FM phase, whose transition
temperature, however, seems to be too low \cite{argyriou99}. Further work
is desirable to clarify this.

This work was supported by a URC fund at HKU.

\end{document}